\documentclass[10pt]{article}
\usepackage[margin=1in]{geometry}

\usepackage{color}
\usepackage{amsmath}
\usepackage{amssymb}
\usepackage{graphicx}
\usepackage{xr}
\usepackage{float}
\usepackage{chemarrow}
\usepackage{multicol}	

\newcommand{\lbda}[4]{\lambda_{#1\to#3}^{#2}}

\newcommand{\comment}[1]{}

\definecolor{WOred}{rgb}{0,0.1,0}

\definecolor{WOred}{rgb}{0.75,0,0}

\definecolor{WOgray}{rgb}{0.5,0.5,0.5}

\definecolor{CGgreen}{rgb}{0,0.7,0}

\begin{document}

\renewcommand{\thefootnote}{\arabic{footnote}}

\title{Transcriptional delay stabilizes bistable gene networks}

\author{Chinmaya Gupta$^{1}$ \and Jos\'e Manuel L\'opez$^{1}$ \and William Ott$^{1}$ \and Kre\v{s}imir Josi\'c$^{1,2,5}$ \and Matthew R.\ Bennett$^{3,4,5,6}$}



\maketitle
\footnotetext[1]{Department of Mathematics, University of Houston, Houston, TX}
\footnotetext[2]{Department of Biology \& Biochemistry, University of Houston, Houston, TX}
\footnotetext[3]{Department of Biochemistry \& Cell Biology, Rice University, Houston, TX}
\footnotetext[4]{Institute of Biosciences \& Bioengineering, Rice University, Houston, TX}
\footnotetext[5]{These authors contributed equally}
\footnotetext[6]{To whom correspondence should be addressed: matthew.bennett@rice.edu}

\begin{abstract}
Transcriptional delay can significantly impact the dynamics of gene networks. Here we examine how such delay affects bistable systems. We investigate several stochastic models of bistable gene networks and find that increasing delay dramatically increases the mean residence times near stable states. To explain this, we introduce a non-Markovian, analytically tractable reduced model. The model shows that stabilization is the consequence of an increased number of failed transitions between stable states. Each of the bistable systems that we simulate behaves in this manner.
\end{abstract}


Transcriptional delay~\footnotemark[7]\footnotetext[7]{Here, we define "transcriptional delay" as the time from initiation of protein production to the first binding of a target promoter. This time includes not only transcription, but also, {\em e.g.}\ translation, folding, and diffusion.} in gene networks is the dynamical consequence of the sequential nature of protein production \cite{mcadams:1995, smolen:2002, mather:2009, josic:2011}. For transcriptional signaling this delay is further compounded by the time it takes for transcription factors to find their target promoters \cite{berg:1981, li:2009}. Previous theoretical work has shown that such delay can significantly affect the dynamics of gene networks and play an important role in a variety of naturally-occurring genetic network architectures.  For example, delay produces oscillations in models of networks containing transcriptional negative feedback loops~\cite{amir:2010, goodwin:1965, lewis:2003, mather:2009, monk:2003, smolen:1999}.  In addition, delayed negative feedback is theorized to govern the dynamics of circadian oscillators~\cite{smolen:2002, sriram:2004}, a hypothesis experimentally verified in mammalian cells~\cite{ukai:2011}.  Experiments also suggest that transcriptional delay can produce robust, tunable oscillations in synthetic gene circuits~\cite{stricker:2008, tigges:2009}.

\begin{figure*}[t]
\centering
\includegraphics[]{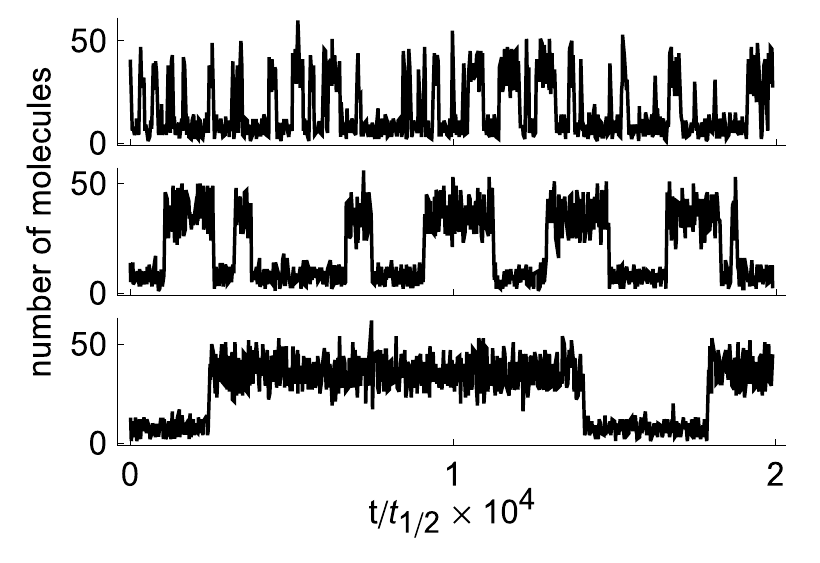}\vspace*{-4mm}
\caption{{\bf Sample trajectories for a single gene positive feedback loop}. From top to bottom, the three timeseries
correspond to transcriptional delays $\tau = 0, t_{1/2}$, and $2t_{1/2}$, where $t_{1/2}$ is the half-life of the protein.}
\label{F:fig1}
\end{figure*}
In this Letter we study the impact of delay on bistable gene networks. Bistability is a central characteristic of biological switches: It is essential in the determination of cell fate in multicellular organisms \cite{hong:2012}, the regulation of cell-cycle oscillations during mitosis \cite{he:2011}, and the maintenance of epigenetic traits in microbes~\cite{ozbudak:2004}. Due to the stochastic nature of gene expression, bistable gene networks can randomly switch between stable states \cite{kepler:2001}. This phenomenon has been studied in many contexts, including the lysis/lysogeny switch of bacteriophage $\lambda$ \cite{aurell:2002, warren:2005}, bacterial persistence \cite{balaban:2004}, 
and synthetically constructed positive feedback loops \cite{gardner:2000, nevozhay:2012}.

Delay and bistability have been extensively studied.  However, the impact of transcriptional delay on bistable gene networks is still unknown. In general, the interaction between delay and stochasticity is complex. Modeling suggests delay affects the stochastic properties of gene expression~\cite{bratsun:2005, gronlund:2010, maithreye:2008, scott:2009}, and that stochastic delay can accelerate signaling in genetic pathways~\cite{josic:2011}. 

\begin{figure*}[tpb!]
\centering
\includegraphics[width=1\linewidth]{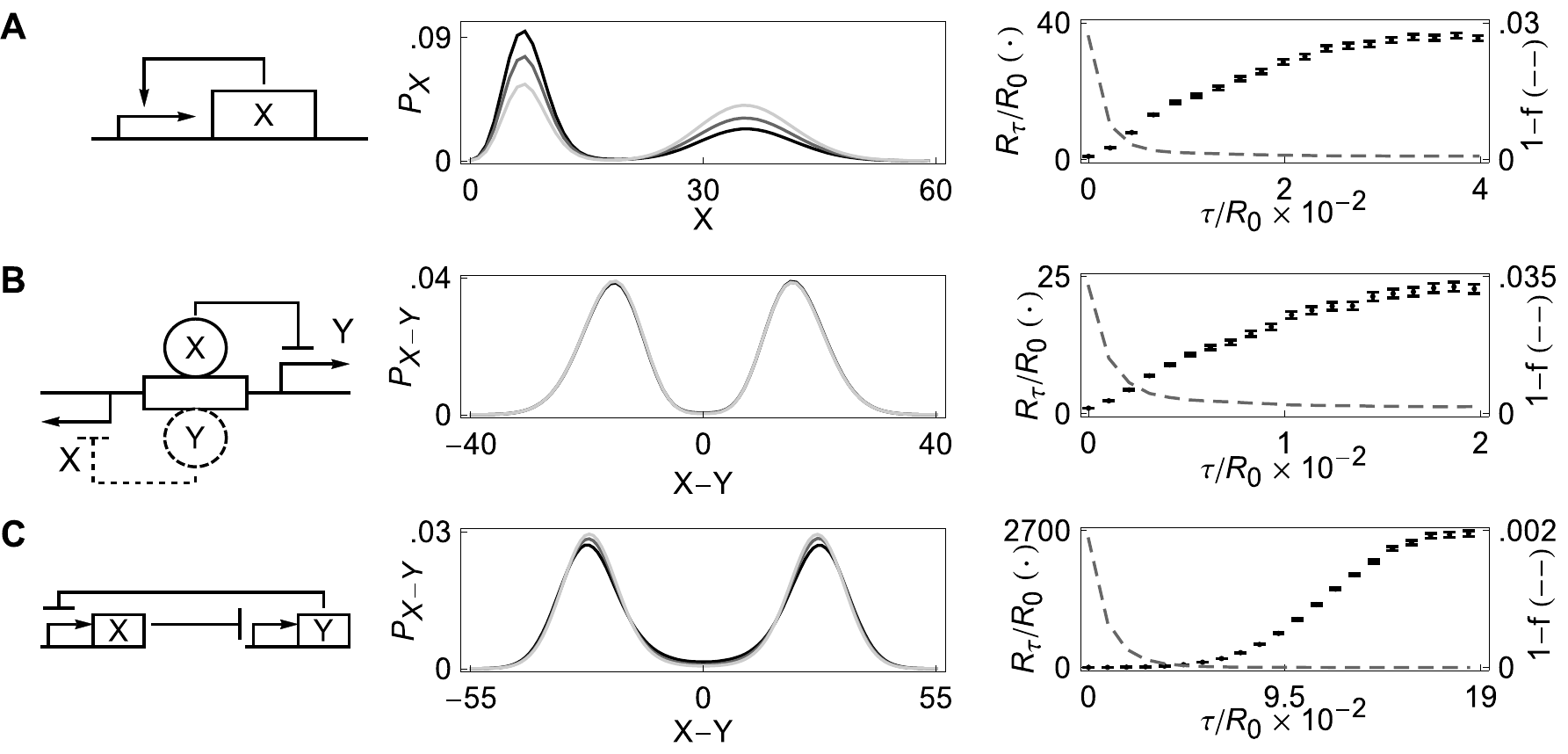}
\caption{{\bf The impact of transcriptional delay on three different
genetic networks.} Left panels show the three different gene networks, 
center panels show the stationary distributions
at three different values of transcriptional delay, $\tau$, and 
right panels illustrate the increase in residence times (dots) and the
decrease in the probability of a successful transition (dashed lines)
with increasing $\tau$.  $R_{\tau }$ denotes the mean residence time in the metastable states at delay $\tau $.  The lighter stationary distributions correspond to larger delays; {\bf (A)} Positive feedback model with stationary distributions at $\tau = 0, 1, 2$, and $R_0 = 227$; {\bf (B)} $\lambda-$phage model with stationary distributions at $\tau = 0,5,10$, and $R_0 = 4829$; {\bf (C)} Co-repressive toggle switch with stationary distributions at $\tau = 0, 0.45, 0.9$, and $R_0 = 489$.}
\label{fig:sim_plots}
\end{figure*}

Here, we investigate a variety of bistable gene networks using a modified version of the Gillespie algorithm that allows us to incorporate transcriptional delay~\cite{schlicht:2008}.  Although the details and dimensionality of the networks differ, in each  the mean residence times near the stable states \emph{increase} dramatically with even modest increases in transcriptional delay time (See Fig.~\ref{F:fig1}).  In some cases, stability increases despite the fact that the stationary distributions show no appreciable change. To explain this phenomenon, we construct a non-Markovian, analytically tractable model. The model predicts that the enhanced stability is due to an increase in the number of failed transitions between stable states.  Stochastic simulations of each bistable system verify this prediction.

{\em Models and simulations.}---To explore the impact of transcriptional delay on  bistable gene networks, we simulated three common systems: 1) A single-gene positive feedback loop; 2) the co-repressive toggle switch \cite{gardner:2000}; and 3) the lysis/lysogeny switch of phage $\lambda$ \cite{warren:2005}. \comment{; and 4) the white/opaque epigenetic switch of {\em C.\ albicans} \cite{lohse:2009}.}
Details about the models, parameters, and simulations can be found in the Supplementary Information (SI).

Consider the single-gene positive feedback loop shown in Fig.\ \ref{fig:sim_plots}-A. The corresponding deterministic dynamics are given by the delay differential equation
\begin{equation}\label{eq:dde_posfeed}
\dot x = \underbrace{\alpha +\beta\frac{x(t-\tau)^b}{c^b+x(t-\tau)^b}}_{\text{birth}}-\underbrace{\gamma x}_{\text{death}},
\end{equation}
where $x$ is the number of proteins per unit volume, $\alpha$  the basal transcription rate due to leakiness of the promoter, $\beta$ the increase in transcription rate due to protein binding to the promoter, $b$ the Hill coefficient, $c$ the concentration of $x$ needed for half-maximal induction, $\gamma$ the degradation rate coefficient of the protein, and $\tau$ is the transcriptional delay time.

This positive feedback loop is bistable for a range of parameters.  As shown in the  SI, the stability of the two fixed points generally \emph{decreases} with an increase in delay in Eq.~\eqref{eq:dde_posfeed}. 

In the corresponding stochastic model, births and deaths occur at rates indicated in the equation.   We chose parameters for which the system switches 
stochastically between the two stable states, and examined a biophysically relevant range of delays.  These were on the order of the protein half-life, but small compared to the transition timescales.

While an increase in delay destabilized the fixed points in the deterministic model, in the stochastic counterpart it resulted in a sharp increase in the  average residence time near the stable states (See Fig.~\ref{F:fig1} and \ref{fig:sim_plots}-A, panel 3). Qualitatively similar behavior was observed in  other models of bistable gene networks we examined (See Fig.~\ref{fig:sim_plots}). Although these systems are quite different, increasing transcriptional delay has qualitatively the same effect in all cases.  

Simulations showed that with an increase in transcriptional delay: a) the mean residence time near the stable states increases (right panels of Fig.~\ref{fig:sim_plots}); b) the probability of a successful transition decreases; and c) the increase in stability may not be accompanied by a consistent change in the stationary distribution (center panels of Fig.~\ref{fig:sim_plots}). These observations appear to be independent of the model system and therefore may have an underlying, unified explanation. However, since the stationary distributions do not {\em necessarily} change in the manner predicted by Kramers' theory \cite{hanggi:1990}, a new explanation of the phenomenon is necessary.

{\em Reduced model.}---In order to obtain a unified description of  the observed increase in stability with an increase in transcriptional delay, we introduce a generalized 3-state reduced model (RM). Two of the states in the model correspond to neighborhoods of the two stable fixed points. We call these states $H$ (high) and $L$ (low). The third state is an intermediate state, $I$, corresponding to a neighborhood of the separatrix. All transitions between $H$ and $L$ must pass through $I$. Therefore, the RM represents a coarse projection of a general bistable model where large fluctuations push the system from the stable states  into a neighborhood of the separatrix.

\begin{figure*}[]
\centering
\includegraphics[width=0.47\linewidth]{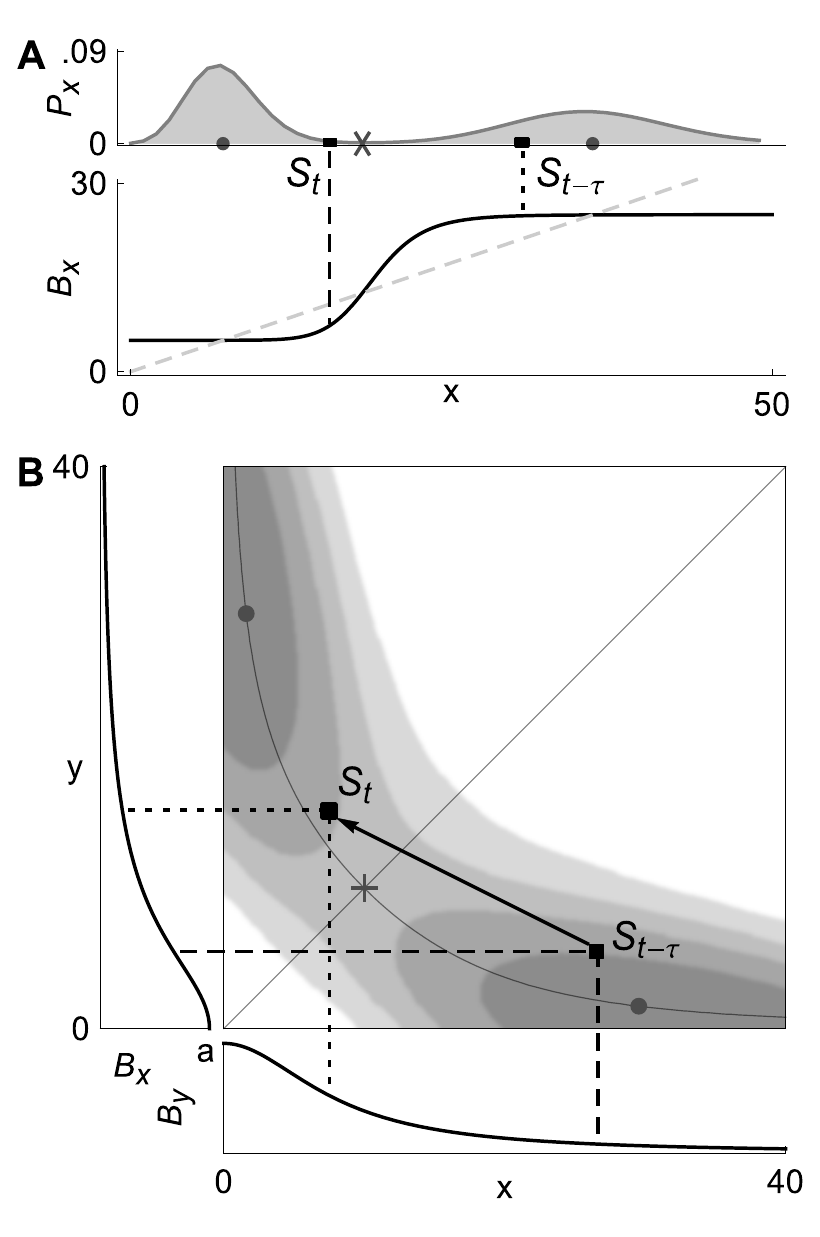}
\vspace*{-0.5cm}
\caption{{\bf Phase space dynamics.} The state of the system at time $t - \tau$ and time $t$ is shown in the case of ({\bf A}) a positive feedback loop, and ({\bf B}) a genetic toggle switch. Mature proteins enter the population at time $t$ at rates determined by the state at time $t - \tau$. In ({\bf A}), at time $t$, the birth rate $B_x$ of $x$  is larger 
in the past, and production is higher than if the birth rate was determined by the present state of the system, $S_t$. This facilitates a return to the
previously visited stable state. A similar explanation holds for ({\bf B}) (see text). Stationary densities are shown in both cases. }
\comment{
}
\label{fig:model_reduction}
\end{figure*}

Due to transcriptional delay, the transition rates between the states  depend on the history of the system. This is particularly important when the system is in state $I$. In the absence of delay the system has no memory; the likelihood of a transition to either stable state from a neighborhood of the separatrix is determined only by the present state of the system. However, in the presence of delay, this likelihood will depend on the past. 

As a particular example, consider the positive feedback loop (Fig.~\ref{fig:sim_plots}A).  At the upper stable state,
which corresponds to state $H$ in the RM, protein production is high.
Consider a large fluctuation from this state that takes the system away from the upper fixed point, to
 state $S_t$ shown in Fig.~\ref{fig:model_reduction}A.  In the presence of transcriptional delay, 
birth rates are determined by the state $S_{t- \tau} = x(t-\tau)$. The larger $\tau$, the more likely it is that $x(t-\tau)$ is in state $H$, near the fixed point whose neighborhood has just been abandoned.  But the birth rates in state $H$ are high and favor motion back toward $H$ (See Fig.~\ref{fig:model_reduction}A).  Therefore, the trajectory is pulled back towards the stable state it came from.
Thus, memory in the system acts as a ``rubber band'' and causes resistance to transitions. 

The situation is similar for the genetic toggle switch, a network of two mutually repressing genes expressing proteins $x$ and $y$.   At the stable states  of this system (dots in Fig.~\ref{fig:model_reduction}B), the
birth rate of one protein is high, and that of the other is low.  Consider a fluctuation away from state $H$, where
$x$ is high and $y$ low, to state $S_t = (x(t), y(t))$.
Birth rates are again determined by the state $S_{t - \tau} = (x(t-\tau), y(t-\tau))$. As shown in Fig.~\ref{fig:model_reduction}B, in state $S_{t - \tau}$ \emph{both} the birth rates of $x$ and $y$ at $t -\tau$ favor motion back to $H$.  Hence
in systems of interacting genes, the ``rubber band'' effect may be even stronger.

To capture the effects of memory in the RM, the transition rates between states are assumed to depend on the state of the system in the past. We define $\lbda{j}{i}{k}{}$, for $i,j,k \in \{H,I,L\}$, as the rate of transition from state $j$ to state $k$, given that   $\tau$ units in the past the system was in state $i$. Not all transitions are possible, as  transitions out of states $H$ and $L$ must go into state $I$. 

We make several assumptions on these transition rates. First, the delay $\tau $ is small compared to the mean residence time in each of the stable states.  Therefore, if the system is in state $H$ or $L$ at time $t$, it is unlikely that  
it was in state $I$ at time $t - \tau$.

Second, we assume that the six transition rates, $\lbda{I}{j}{k}{},$ out of state $I$ are at least an order of magnitude larger than transition rates out of states $L$ and $H$. This corresponds to the assumption that the system
will exit the vicinity of the separatrix much more quickly than the vicinity of a stable fixed point.

Finally, we assume that for time $\tau$ after entering state $I$, the system is more likely to return to its previous state. In other words, we assume
\begin{equation}
\label{eq:bias}
p^H_{I\to H} > p^I_{I\to H}, \qquad
p^{L}_{I \to L} > p^I_{I \to L},
\end{equation}
where
\begin{equation*}
p^{i}_{I \to j} = \frac{\lbda{I}{i}{j}{}}{\lbda{I}{i}{H}{} +\lbda{I}{i}{L}{}}
\end{equation*}
is the probability of transitioning from state $I$ to state $j\in\{H,L\}$ given the system was in state $i$ a time $\tau$ in the past. This assumption captures the ``rubber band'' effect illustrated in Fig.\ \ref{fig:model_reduction}, {\em i.e.}\ delayed protein production favors a return to the stable state that was visited last.

{\em Metastability as a function of delay.}---We now analyze the stability of states $H$ and $L$ as a function of the delay, $\tau $.  Let $R_{H}$ denote the residence time for state $H$.  We compute the expected value $E[R_{H}]$; the computations for $L$ are analogous.  Once the system enters state $H$, it will make a number of failed transitions of the form $H \to I \to H$ before eventually making a successful transition $H \to I \to L$.  Let $f_{H}$ denote the probability of a failed transition, that is, the probability that a transition $H \to I \to H$ occurs conditioned on the system having been in state $H$ for at least time $\tau $.  We have
\begin{equation}\label{eq:prob_fail}
f_H = (1 - Z_H(\tau)) p^{H}_{I\to H} + Z_H(\tau) p^I_{I \to H},
\end{equation}
where we define 
$Z_{H} (\tau )$ by 
\begin{equation*}
Z_H(\tau ):= \exp( - (\lbda{I}{H}{H}{} + \lbda{I}{H}{L}{}) \tau ).
\end{equation*}
Note that $f_{H}$ is a convex linear combination of $p^{H}_{I \to H}$ and $p^{I}_{I \to H}$.  When $\tau = 0$ (Markovian case), $f_{H} = p^{I}_{I \to H}$. As $\tau $ increases away from zero, $f_{H}$ moves toward $p^{H}_{I \to H}$.

Let $F_{H}$ denote the random time needed to complete a failed transition and let $S_{H}$ denote the time needed for a successful transition.  Assuming that the delay is small compared to the characteristic residence times in the stable states, we obtain the key estimate for $E[R_{H}]$.  Writing $R = R_{H}$, $f = f_{H}$, $F = F_{H}$, $S = S_{H}$, and $\lambda = \lbda{H}{H}{I}{}$, we have
\begin{equation}
\label{e:mean_residence_time}
E[R] \sim \frac{f}{1-f} \left( E[F] + \frac{1}{\lambda } \right) + E[S] + \frac{1}{\lambda }.
\end{equation}

The primary contribution in Eq.~\eqref{e:mean_residence_time} comes from the term $f(1 - f)^{-1}$ representing the mean number of failed transitions of the form $H\to I \to H$ before a successful transition $H \to I \to L$. The terms $E[F]$ and $E[S]$ are not very sensitive to $\tau$. On the other hand, because of the inequalities~\eqref{eq:bias}, $f(1-f)^{-1}$ grows rapidly as $\tau $ increases from $\tau = 0$ . If the states $H$ and $L$ are sufficiently stable, the expected time spent in each state before a large fluctuation is approximately $\lambda^{-1}$. 

\begin{figure*}[tp!]
\centering
\includegraphics[width=0.47\linewidth]{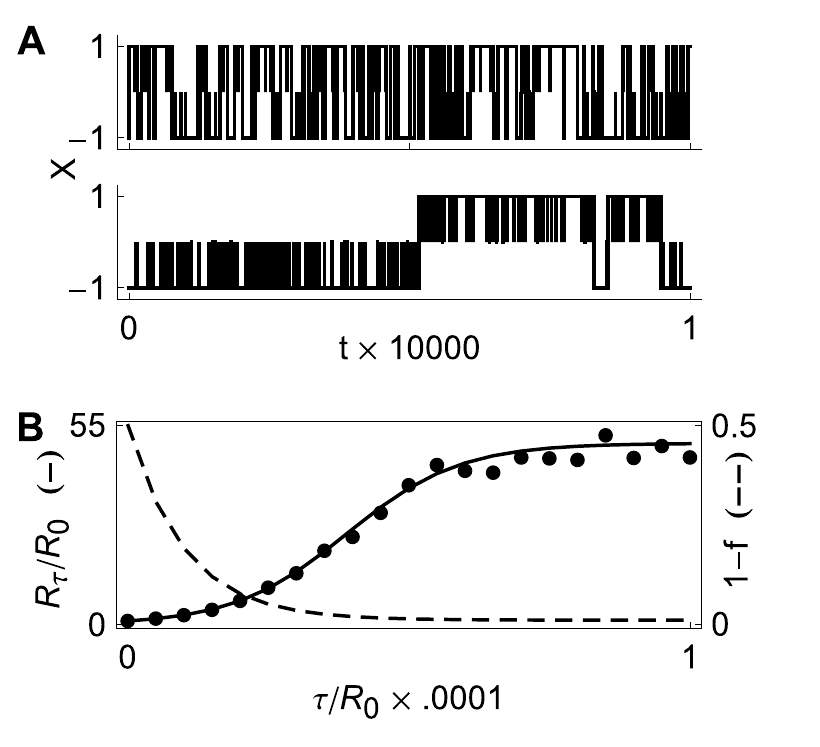}

\caption{ {\bf The reduced model}
({\bf A}) Sample trajectories for the 3-state RM. Top and bottom trajectories correspond to $\tau = 0$ and $\tau = 0.03$, respectively. ({\bf B}) The estimated increase in mean residence time as a function of delay given by Eq.~\eqref{e:mean_residence_time} \emph{(Left axis, solid line)} compared to values obtained by simulating the RM (bold circles). 
The dashed line represents the probability of a successful transition as a function of delay \emph{(Right axis)}.
}

\label{fig:sim_match_theory}
\end{figure*}

Fig.~\ref{fig:sim_match_theory}-B illustrates that the RM qualitatively behaves like the models shown in Fig.~\ref{fig:sim_plots}.

{\em Concluding remarks.}---The existence of multiple stable states is a common feature of genetic networks, such as those that determine cell fate in multicellular organisms \cite{hong:2012}. Since noise is ubiquitous in gene networks,  mechanisms that stabilize dynamics are essential. We have shown that transcriptional delay can stabilize bistable gene networks. The rates at which mature proteins enter the system depend on its past state, and the tendency to return to the state from which the system just escaped increases with delay. 

The RM proposed in this Letter depends on neither the explicit underlying model nor the specific distribution of delay times. Hence, we predict that distributed delay will stabilize bistable systems, provided the delay distribution is not close to  the residence times of the system.  As shown in SI Fig.~\ref{fig:distDelay}, this is indeed the case. 

The RM can also be modified to explain a \emph{decrease} in stability if degradation is delayed and production is instantaneous (see SI Fig. \ref{fig:delay_death}).  The explanation given in Fig.~\ref{fig:model_reduction} carries over
to this case:  Delayed death pushes the system away from the stable state that was just visited.

In previous studies of bistable systems in the presence of delay, it was assumed that the delay time was on the order of the residence time~\cite{Tsimring2001, Masoller2003}.  Since transitions are relatively fast compared to this timescale, it was sufficient to consider reductions with only two states, $H$ and $L$.  However, transcriptional delays in gene networks are typically much smaller.  As we have shown here, in such systems it is therefore appropriate to introduce an additional intermediate state, $I$, in a reduced model.

Transcriptional delay introduces a number of other dynamical changes to stochastic bistable systems. For instance, typical transition paths between the stable states change with delay (see SI Fig.~\ref{fig:densities}).  Such  changes depend on the specifics of the individual models, and  cannot be understood using the reduction described above. 
 Schwartz {\em et al.}\ examined stability and most probable transition paths for Langevin equations with delay-dependent drift and non-delayed diffusion~\cite{Schwartz}.  An extension of their approach may allow a more detailed
 analytical examination of the systems described here.


{\em Acknowledgments}---We thank Lee DeVille for insightful discussions.  This work was funded by the NIH, through the joint NSF/NIGMS Mathematical Biology Program grant R01GM104974, Robert A. Welch Foundation grant C-1729, and a John S. Dunn Foundation Collaborative Research Award Program administered by the Gulf Coast Consortia.

\bibliographystyle{plain}
\bibliography{AllPapers_v5,mattsliterature}


\renewcommand{\lbda}[4]{\lambda_{#1\to#3}^{#2}}
\newcommand{\Lbda}[4]{\Lambda_{#1\to#3}^{#2}}
\newcommand{\ALPHA}{a}

\renewcommand{\theequation}{S\arabic{equation}}
\renewcommand{\thefigure}{S\arabic{figure}}

\newcommand{\hilite}[1]{{\color{red} #1}}

\renewcommand{\comment}[1]{}

\newpage

\section*{Supplimentary Material}

\paragraph{Outline.} We present here the details of several computations that are described in the main manuscript. We also describe in detail all the models that have been used in the manuscript, and expand on various notes in the discussion in the manuscript. 

 First, we present the detailed analysis of the reduced model, and derive the primary expression of interest. Second, we outline the modifications to the stochastic simulation algorithm required to incorporate delay (page \pageref{sec:dSSA}). Third, we provide a discussion of the positive feedback model, and present the parameters used for simulation (page \pageref{sec:PF}), as well as discuss the changes in the stability of the fixed points for the deterministic approximation (page \pageref{sec:PF-stability}). We also discuss the case of distributed delays (page \pageref{sec:PF-distributed}), and delayed deaths (page \pageref{sec:PF-delay-death}). Fourth, we present the details of the lysis/lysogeny switch of phage $\lambda$ (page \pageref{sec:LL}). Finally, we present the details of the co-repressive toggle switch, along with a discussion of a geometric method for finding parameters for which the system exhibits stochastic transitions between the stable states (page \pageref{sec:CRT}). 

In all models time is scaled so that one unit of time corresponds to the half-life of a protein.

\section{Analysis of the reduced model (RM)} \label{sec:RM}

We recall here the structure of the RM and derive the various estimates of interest.  The states $L$ and $H$ in the  RM correspond to the those regions of the phase space that are in the vicinity of a stable point. While the process is in the
potential well, it undergoes small fluctuations around the fixed points of the corresponding deterministic system. 

The third state, $I$, is an intermediary state. All transitions from the basin of one stable point must cross through the intermediary state before they can fall into the second basin. In the phase space, the intermediary state is represented by a 
thick neighborhood of the separatrix for the basins of attraction.
As described in the main manuscript, since the system retains memory, the immediate history is an important consideration around the separatrix. A fluctuation may push the current state of the system from one basin to another, but mature molecules entering the population can push the system back into the old potential well. This is markedly different from the Markovian case, where only the current population composition is needed for determining future probabilities. 

The idea behind the analysis of the RM is to first discretize time using a step size $\Delta$ and to obtain the probability of making failed transitions for a given delay $\tau$. The probability density function for the random variable that counts the number of failed transitions in the continuous limit is obtained by taking the limit $\Delta\to 0$. This, in turn, allows us to compute the mean number of failed transitions and the mean time spent in a failed transition, as well as the mean time spent in a successful transition, assuming that the residence times dominate the delay. These are the primary ingredients required for the computation of the residence times in the stable states.

We study the discrete case first. If the delay is assumed to be $K\Delta$ (where $K$ is some positive integer), the RM can be embedded in a $(K+1)$-dimensional space, and can be represented by vectors $\vec{x} = (x_{-K\Delta}, x_{-(K-1)\Delta}, \dots, x_{-\Delta}, x_0)$ with the state $x_0$ being the current state of the RM and $x_{-t\Delta}$ being the state $t$ steps in the past. In the higher-dimensional space, not all jumps are feasible, and the process can only possibly jump from a configuration $\vec{x}$ to a configuration $\vec{y}$ if $x_{-i \Delta} = y_{-(i+1) \Delta}$ for all $0 \leq i \leq K-1$. The delay can now be expressed by saying that the probability of a feasible jump from $\vec{x}$ to $\vec{y}$ is given by $\Lbda{x_0}{x_{-K\Delta}}{y_0}{}$. Call $f_{H}$ the probability of failing a transition, given that the transition starts in the state $H$, and $x_{-i\Delta} = H$ for all $0 \leq i \leq K$.  Let $P(k)$ denote the probability mass function for the number of steps $k$ that are required to complete the loop $H \to I \to H$ given that $H \to I$ has occurred. We see that
\begin{equation}\label{pdf-failed-discrete}
P(k) = \frac{1}{f_{H}}\begin{cases}
(\Lbda{I}{H}{I}{})^{k-1} (1 - \Lbda{I}{H}{I}{}) \frac{\Lbda{I}{H}{H}{}}{\Lbda{I}{H}{H}{}+\Lbda{I}{H}{L}{}} & 1\leq k \leq K;\\
(\Lbda{I}{H}{I}{})^K (\Lbda{I}{I}{I}{})^{k - K -1}(1 - \Lbda{I}{I}{I}{}) \frac{\Lbda{I}{I}{H}{}}{\Lbda{I}{I}{H}{} + \Lbda{I}{I}{L}{}} & k \geq K+1.
\end{cases}
\end{equation}

In the continuous-time limit, a discrete delay $K\Delta$ is replaced by a delay $\tau$ where $\Delta\to 0$ and $K\Delta \to \tau $.  As in the discrete case, for the continuous-time system let $f_{H}$ denote the probability of failing a transition, given that the transition initiates from state $H$ and the process remembers only state $H$ when the transition begins.  Let $P(t)$ denote the probability density function for the random variable $F_{H}$: the time needed to complete the $H \to I \to H$ loop given that $H \to I$ has occurred.

\begin{table}
\centering
\begin{tabular}{|c|c|c|c|}
\hline
 & {\bf H} & {\bf I} & {\bf L}\\
 \hline\hline
 $\tau = 0$ & 0.502 & $2.019 \times 10^{-5}$ & 0.498\\
 $\tau = 0.04$ & 0.498 & $1.977 \times 10^{-5}$ & 0.502\\
 $\tau = 0.08$ & 0.500 & $2.045 \times 10^{-5}$ & 0.499\\
 \hline
\end{tabular}
\caption{Shown here are the stationary distributions for the RM for three different values of $\tau$: $\tau = 0, \tau = 0.04$ and $\tau = 0.08$. Rates used in the RM are $\lambda^{I}_{I\to x} = 50, \lambda^{x}_{I\to x} = 99, \lambda^{x}_{I\to x^c} = 1,   \lambda^{I}_{x \to I} = 0.20, \lambda^{x}_{x \to I} = 0.0002, \lambda^{x}_{x^c \to I} = 0.004; x\in\{H,L\}; x^c = H \text{ if } x = L \text{ and } x^c = L \text{ if } x = H$.}
\end{table}

For the continuous-time case, we compute a formal limit in Eq.~\eqref{pdf-failed-discrete}. To do so, we set up some notation. For a continuous-time Markov process, a transition probability in a time interval of length $\Delta$ is given by $\lambda \Delta$ were $\lambda$ is the corresponding transition rate for the process. In the discrete-time description of the process, we can replace probabilities such as  $\Lbda{i}{j}{k}{}$ ($i \neq k$) by $\lbda{i}{j}{k}{} \Delta$ (rates corresponding to transitions) and probabilities $\Lbda{I}{j}{I}{}$ by $(1 - (\lbda{I}{j}{H}{} + \lbda{I}{j}{L}{})\Delta )$ for $j \in \left\{ H, I, L \right\}$.  For fixed $t$ and any $\Delta $ such that $t \Delta^{-1}$ is a positive integer, Eq.~\eqref{pdf-failed-discrete} gives
\begin{equation}
 P(F_{H} \in [t - \Delta, t])=  \frac{1}{f_{H}}\begin{cases}
 \left(1 - \left(\lbda{I}{H}{H}{} + \lbda{I}{H}{L}{}\right)\Delta\right)^{t \Delta^{-1} -1} \lbda{I}{H}{H}{} \Delta & \Delta \leq t \leq \tau\\
 \left( 1 - \left( \lbda{I}{H}{H}{} + \lbda{I}{H}{L}{}\right) \Delta\right)^{\tau \Delta^{-1}} \left( 1 - \left( \lbda{I}{I}{H}{} + \lbda{I}{I}{L}{}\right)\Delta\right)^{(t - \tau)\Delta^{-1} -1} \lbda{I}{I}{H}{} \Delta & t \geq \tau + \Delta.
 \end{cases}
\end{equation}
Taking $\Delta\to 0$, we obtain
\begin{equation}\label{pdf-failed-continuous}
 P(t)  = \frac{1}{f_{H}} \begin{cases}
 \lbda{I}{H}{H}{} \exp \left( -(\lbda{I}{H}{H}{} + \lbda{I}{H}{L}{}) t\right) & 0 < t \leq \tau\\
 \lbda{I}{I}{H}{} \exp \left( -(\lbda{I}{H}{H}{} + \lbda{I}{H}{L}{}) \tau - (\lbda{I}{I}{H}{} + \lbda{I}{I}{L}{})(t - \tau)\right) & t > \tau
 \end{cases}.
\end{equation}
Since $P(t)$ is a pdf, it follows from integrating on $[0, \infty )$ that 
\[
f_{H} = (1 - Z_{H} (\tau )) p_{I\to H}^H \;\;+\;\;  Z_{H} (\tau ) p_{I\to H}^I
\]
where
\[
p^{i}_{I\to H} = \frac{\lbda{I}{i}{H}{}}{\lbda{I}{i}{L}{} + \lbda{I}{i}{H}{}}, \quad Z_{H} (\tau ) := \exp(-(\lbda{I}{H}{H}{} + \lbda{I}{H}{L}{})\tau).
\] 

Analogous computations can be performed for the probability of failing a transition if it is assumed that the transition initiates from state $L$, and the only remembered state is $L$. Analogous computations can also be performed to obtain the probability density function for the time spent in making a successful transition loop $H \to I \to L$, and, therefore, the probability of a successful transition. Further, we can compute the expectations of $F_{H}$, and $S_{H}$ (defined as the time spent in a successful transition):
\[
E[F_{H}] = \int_{0}^{\infty} t P(t) \; dt
\]
et cetera.  The exact forms of these expressions can be computed analytically, but we do not write them here. Instead, we note that $d E[F_{H}]/d \tau $ and $d E[S_{H}]/d \tau $ are very small. This implies that the times spent in state $I$ during a failed transition, or a successful one, do not significantly change with $\tau$. 

Denote by $N$ the number of failed transitions before a successful transition. Then $N$ has a geometric distribution ($P(N = n) = f_{H}^n (1-f_{H}), n\geq 0$).

In between each failed transition, the process spends time in the stable states $H$ or $L$ before jumping out on another excursion. We have assumed that the residence times sufficiently dominate the delay; a consequence of this assumption is that once the process re-enters the state $H$, it stays there long enough to forget its past excursions. Therefore, we can estimate the time between transition attempts by $1/\lambda$ where $\lambda = \lbda{H}{H}{I}{}$.

We can now estimate the expected residence time in the stable state:
\[
E[R_{H}] \sim \frac{f_{H}}{1-f_{H}}\left(
E[F_{H}] + \frac{1}{\lambda}
 \right) + E[S_{H}]  + \frac{1}{\lambda}.
\]

\section{Gillespie's Stochastic Simulation Algorithm with Delay}\label{sec:dSSA}

Gillespie's stochastic simulation algorithm (SSA) is a way to sample exact stochastic realizations of a chemical system. In the SSA, the reactions are modeled as birth-death processes.  In the classical SSA at time $t$ the state of the system is described by the population vector ${\bf x}(t) = (x_1(t), \dots, x_N(t))$. The set of possible reactions in the system is indexed by $\left\{ 1, 2, \dots, M\right\}.$ For each reaction $j$, a propensity function $a_j: \mathbb{R}^N\to\mathbb{R}^+$ describes the rate at which the reaction fires, given the population ${\bf x}(t)$. A vector ${\bf v}_j \in \mathbb{Z}^N$ describes the change in each species when reaction $j$ fires. The total rate of all reactions together is given by $\sum_j a_j(x(t)).$ The SSA proceeds by first sampling a time, $\Delta$, to the next reaction from the exponential distribution with mean $\sum_j a_j$.  The reaction is assigned type $i$ with probability  $a_i /\sum_j a_j$. The population ${\bf x}(t)$ is then updated by adding the appropriate state change vector ${\bf x}(t)\mapsto{\bf x}(t)+{\bf v}_i.$

In the examples considered in the manuscript some reactions (in most of the manuscript these are births) only affect the population size after a delay, while others (in most of the manuscript these are deaths) affect the population size immediately.  Delays need not be of constant length.  In the case of non-constant delays we assume that they are i.i.d.  with distribution $\kappa(\tau)$.  

To simulate such processes we used a modified version of Gillespie's algorithm~\cite{schlicht:2008}: At time $t$ in the simulation the time to the next reaction, $\Delta$, and the type of reaction, $i,$ is sampled, as described above. Before proceeding $\Delta$ units of time, one checks to see if there are any reactions that commenced in the past, and finish in the interval $[t, t+\Delta]$. If there are no such reactions, one proceeds to time $t+\Delta$. If the reaction sampled, $i$, is a non-delayed reaction, the population is updated immediately. If $i$ is a delayed reaction, the state change vector corresponding to $i$ is put in a queue, with a designated time of exit $\tau$ units of time in the future, sampled from some distribution $\kappa(\tau)$. 

If on the other hand, there is a reaction from the past which terminates in the time $[t, t+\Delta]$, one proceeds to the  time of this reaction. The population size is changed according to the state change vector of this past reaction. The original waiting time $\Delta$, and reaction type $i$, are discarded, and a new waiting time and reaction type are sampled.

\section{Positive Feedback Model}\label{sec:PF}

The deterministic delay-differential equation that approximates the stochastic dynamics of the single gene positive feedback model is given by
\begin{equation} \label{eq:PF}
\dot x = \alpha +\beta\frac{x(t-\tau)^b}{c^b+x(t-\tau)^b}-\gamma x
\end{equation}

The parameters used were $\beta = 20$, $\alpha = 5$, $c = 19$, $\gamma = \ln(2)$ and $b = 10$. The parameter $\alpha$ is the basal rate of production of the molecules of $x$ (with units molecules $s^{-1}$). The maximal rate of production for the system is $\alpha + \beta$. $c$ is the number of molecules of $x$ required to achieve the half-maximal activation rate of $\alpha + \beta/2$. The constant $b$ is the Hill coefficient for the activation function, and $\gamma$ is the rate constant for the degradation of the protein $x$ (with units $s^{-1}$).

For this set of parameters, there are three fixed points for the delayed differential equation: $x = 7.2, x = 18.0$ and $x = 36.0$. The fixed points at $7.2$ and $36.0$ are stable, while the fixed point at $x = 18.0$ is unstable. 

We map the phase space of the positive feedback model onto the RM using  $H = [23, \infty)$, $L = [0, 13]$ and $I = (13, 23)$. A trajectory that starts in the state $H$, and makes an excursion into $I$ is said to have a successful transition if it reaches state $L$ before state $H$.

All trajectories are initialized in the state $H$ at $x = 25$, following which a long transient is computed. Any data is gathered after the transient. The mean residence times $R_\tau$ are computed by averaging over $10^4$ transitions. 
\vspace{0.5cm}

\noindent\paragraph{Stability analysis for the Positive Feedback Model.}\label{sec:PF-stability}

We analyze  the spectrum of the linearization around the fixed point of the delay differential equation~\eqref{eq:PF} rewritten as
\begin{equation} \label{E:DDE}
\dot x(t) = B(x(t - \tau)) - D(x).
\end{equation}
This equation is the deterministic counterpart of the stochastic positive feedback model 
examined in the manuscript. 
Linearizing Eq.~\eqref{E:DDE} in the neighborhoods of a stable fixed point
$x_0$ to yield a DDE
\[
\dot x(t) = B'(x_0) (x(t - \tau) - x_0) + D'(x_0) (x(t) - x_0). 
\] On setting $y(t) = x(t) - x_0$, $p = -B'(x_0)$, $q = D'(x_0)$, and assuming a solution of the form
\[
y(t) = C e^{st}
\] we get a characteristic equation
\[
(s+q)e^{s\tau} + p =  0.
\] The $k^{\mathrm{th}}$pair of eigenvalues $s_k$ can be obtained by solving the equation
\[
s_k = \frac{1}{\tau} W_k(-p \tau e^{\tau q}) - q
\] where $W_k$ is the $k^{\mathrm{th}}$ branch of the Lambert $W$ function. If $s_0$ is found to have
negative real part, no other eigenvalues need to be computed as the stability is 
determined completely by $s_0$.

\begin{figure}[h]
\includegraphics[]{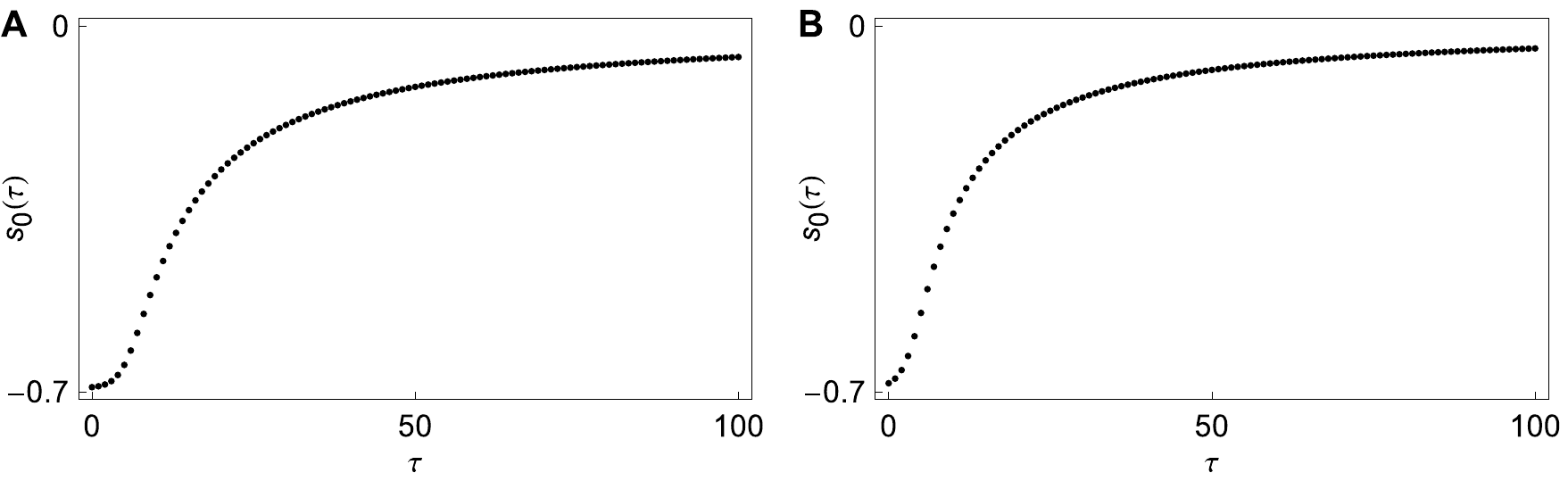}
\caption{{\bf Left: }Plot of the leading eigenvalue, as a function of the delay, for the linearization of the system in the neighborhood of the lower fixed point $x_0 = 7.21$. {\bf Right: }Plot of the leading eigenvalue, as a function of the delay, for the linearization of the system in the neighborhood of the higher fixed point $x_0 = 36.01$.}\label{fig:lin-stab}
\end{figure}

As is apparent from Fig \ref{fig:lin-stab}, as $\tau$ increases, the stability of both stable fixed points
decreases. Therefore, while the stochastic positive feedback system becomes 
mores stable, its deterministic counterpart becomes less stable with an increase in delay.

\vspace*{1cm}

\noindent\paragraph{Distributed Delays.}\label{sec:PF-distributed}

To examine the effect of  distributed delay, we used Gamma delay distributions, $\kappa(\tau; \mu, \sigma),$ with different means and variances. The effect of increasing the mean of the distribution $\kappa$ was qualitatively similar to increasing the
magnitude of a constant delay, $\tau$:  The mean residence times increased sharply with small increases in this mean, and eventually appeared to saturate.

\begin{figure}[h]
\includegraphics[]{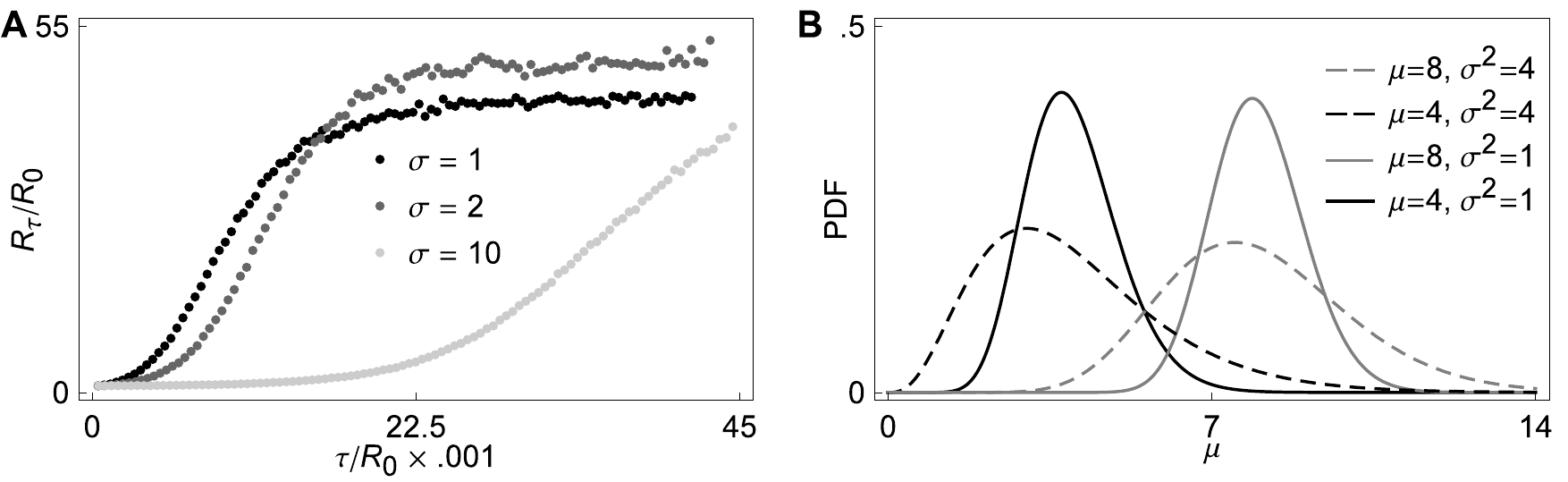}
\caption{{\bf Left: Positive feedback model with distributed delay.} A Gamma distribution with $\tau \in(0, 10)$ and $\sigma \in \left\{1, 2,10\right\}$ is used for the positive feedback model. The parameters used are $a = 20, b = 5, k = 19$. The trajectories are initialized with a population of size 25, followed by a long transient. For small values of $\tau$, the ratio $R_\tau/R_0$ grows more slowly for larger variances. {\bf Right: Delay distributions.} We use families of Gamma distributions with varying means and variances $\sigma^2 = 1, 4$. Displayed here are the distributions for means $4$ and $8$. }\label{fig:distDelay}
\end{figure}

Increasing the variance of the gamma distribution for a fixed mean appears to initially slow down the rate of increase of the mean transition time with increasing delay; however, larger variances also appear to correspond to larger saturation values (see Fig.~\ref{fig:distDelay}). 

\vspace*{1cm}

\paragraph{Delayed Deaths.}\label{sec:PF-delay-death}

\begin{figure}[h]
\includegraphics[width=7in]{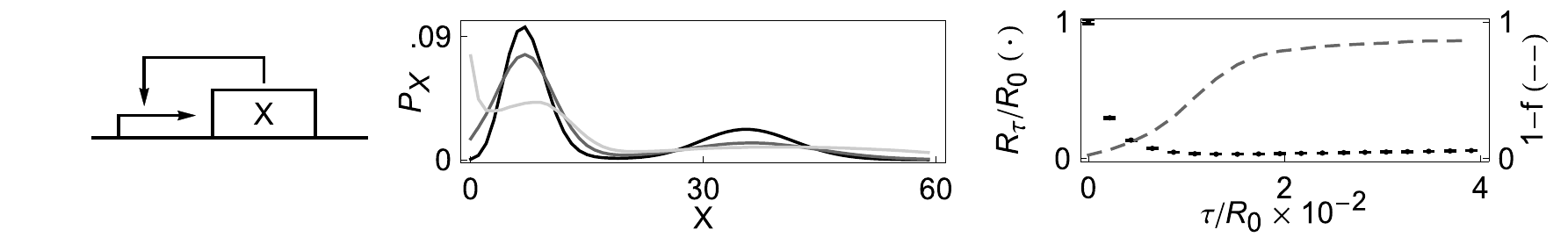}
\caption{{\bf Effect of delaying degradation.} \emph{(Left)} The motif for the positive feedback model. \emph{(Center)} The stationary distributions for three different delays. Darker lines correspond to larger delays. The maximum accumulation of proteins of type $X$ increases with delay. Since the exiting state change vectors depend on the protein numbers at a time in the past, there is an accumulation of probability at $X = 0$ with increasing delay. \emph{(Right)} Solid dots represent the mean residence times. Dashed lines represent the probability of succeeding in an attempted transition.}\label{fig:delay_death}
\end{figure}

We also consider a process that is formally constructed by delaying deaths (see Fig. \ref{fig:delay_death}). In such a model, as in the stochastic simulation algorithm with delayed births, the waiting time to the next reaction is sampled. If that reaction is a birth type reaction, the population is updated immediately. For a degradation, we put the corresponding state change vector in a queue with a designated time of exit. A new reaction time is then sampled. Otherwise the algorithm is as described above.

Delaying reactions that decrease population is less realistic, and can lead to negative population sizes. We disregard any reactions that would decrease the size of a population below 0. Formally, the deterministic equations approximating the stochastic dynamics in the case of delayed deaths can be written as
\begin{equation}
\dot x = \alpha +\beta\frac{x(t)^b}{c^b+x(t)^b}-\gamma x(t - \tau)
\end{equation}
with the added constraint that $x(t) \geq 0$ for all $t\geq 0$.

Delaying deaths destabilizes the bistable switch.  The explanation parallels that given in the main manuscript:Consider a large downward fluctuations from away from the upper stable state $H$.  In the presence of transcriptional delay, 
birth rates are determined by the state $x(t-\tau)$, and the larger $\tau$, the more likely it is that $x(t-\tau)$ is in state $H$.  But death rates in state $H$ are high and favor motion away from $H$.  Therefore, the trajectory is pushed away from the stable state it came from.  The result is a large decrease in residence times with increasing delay.

The RM described in the manuscript can capture this effect by appropriately changing the transition rates $\lambda^{H}_{I\to H}$ etc.

\section{The lysis/lysogeny switch of phage $\lambda$.} \label{sec:LL}

The lysis/lysogeny switch of phage-$\lambda$ is realized as a set of chemical reactions involving multimerized forms of two transcription factors $A$, and $B$, which regulate gene expression by binding to the genome at an operator site $O$. The state of the operator is denoted by $O$ if neither multimerized transcription factors are bound to it; when the multimer $A_n$ is bound, the operator is denoted $OA_n$ and when $B_m$ is bound, the operator is denoted by $OB_m$. 

A simplified model of the system consists of the following chemical reactions. The first two reactions represent the multimerization reactions of the transcription factors $A$ and $B$. Multimerization is introduced into the model because transcription factors must bind to the DNA cooperatively to make a working switch. 

\newcommand{\rl}[2]{\autorightleftharpoons{\ensuremath{#1}}{\ensuremath{#2}}}

\begin{subequations}
\begin{align}
nA \rl{k_{f}}{k_{b}} A_n\\
m B \rl{k_f}{k_b} B_m
\end{align}
\end{subequations}

The next two reactions represent the degradation of the transcription factor monomers as a first order reaction.

\begin{subequations}
\begin{align}
A \autorightarrow{$\mu_A$}{} \emptyset\\
B \autorightarrow{$\mu_B$}{} \emptyset
\end{align}
\end{subequations}

When no multimers are bound to the operator $O$, either $A$ or $B$ can be produced. This production is a result of a large number of biochemical steps, and for this reason, this reaction is assumed to be delayed. This delay is represented in the equations by setting $\tau$ as a superscript in the reaction rate constant. The synthesis reactions for $A$ and $B$ are then represented as
\begin{subequations}
\begin{align}
O \autorightarrow{$k_A^{(\tau)}$}{ } O + A\\
O \autorightarrow{$k_B^{(\tau)}$}{ } O + B.
\end{align}
\end{subequations}

The multimers $A_n$ or $B_m$ can bind reversibly to the operator $O$. This gives us
\begin{subequations}
\begin{align}
O + A_n \rl{k_{on}}{k_{off} } OA_n\\
O + B_m \rl{k_{on}}{k_{off} } OB_m.
\end{align}
\end{subequations}

Once the transcription factors of a certain kind bind to the operator, they allow only for the production of monomers of their own kind. This is represented by
\begin{subequations}
\begin{align}
OA_n \autorightarrow{$k_A^{(\tau)}$}{ } OA_n + A\\
OB_m \autorightarrow{$k_B^{(\tau)}$}{ } OB_m + B.
\end{align}
\end{subequations}

Together, this gives us a complete description of the lysis/lysogeny switch of phage$-\lambda$. We used the parameters $k_b = 5 = k_f = k_{on}, k_{off} = 1, k_A = 1 = k_B, \mu_A= 0.3 = \mu_B$ and $n = m = 2.$  We assume the presence of only 1 operator $O$. A detailed analysis of the system, as well as a discussion of the model reduction of the deterministic approximation to a system of two ordinary differential equations can be found in \cite{Warren2005}.

The phase space of the lysis/lysogeny switch is mapped onto the RM as follows. We denote by $N_X$ the number of molecules of type $X$ in the system, we compute $|A| = N_A + 2 N_{A_2}$ and $|B| = N_B + 2 N_{B_2}$. If $|A| - |B| \geq 5$, the system is considered to be in state $H$, and if $|A| -|B| \leq 5$, in state $L$. Otherwise, if $|A | - |B| \in (-5,5)$, the system is said to be in state $I$. The system is initialized with $5$ molecules of the protein $A$ and $30$ molecules of $A_2$ (making $|A| = 65$) and $|B| = 0$. A long transient is computed before any data is gathered. Mean residence times are computed by averaging over $10^4$ transitions.

\section{Co-repressive toggle switch.}\label{sec:CRT}

 We first describe the deterministic system in order to obtain the critical points. An appropriate choice of parameters is important, since transitions are very rare between stable points which are widely separated. We present here a geometrical method to easily find feasible parameters.

The standard form of the co-repressive toggle switch with delayed production is given by the set of delay ordinary differential equations \cite{gardner:2000}
\begin{align*}
\dot{x} &= \frac{\beta}{1+ y(t - \tau)^2/k} - \gamma x \\
\dot{y} &= \frac{\beta}{1+ x(t - \tau)^2/k} - \gamma y.
\end{align*}
The production and degradation propensity functions for the delayed Gillespie algorithm are obtained from these ODEs. Since the fixed points of the system do not change if delay is introduced, we assume that $\tau = 0$, and parametrize the critical points of the non-delayed system: Setting $\dot{x} = \dot{y} = 0$ and writing $\ALPHA = \beta/2\gamma$ we get
\begin{align*}
0 &= \frac{\beta}{1+y^2/k} - \gamma x = \frac{2\ALPHA k}{k+y^2}-x\\ 
0 &= \frac{\beta}{1+x^2/k} - \gamma y = \frac{2\ALPHA k}{k+x^2}-y. 
\end{align*} We now eliminate $y$ between the two equations to obtain
\[
\frac{\left(2 \ALPHA k - k x - x^3 \right) \left(k+x^2-2 x \ALPHA \right)}{\left(k+x^2\right)^2+4 k \ALPHA ^2} = 0.\]

The numerator is a polynomial of degree 5, which implies that there exist at most 5 real critical points. Solving the quadratic part of the above equations, and substituting back into the original equation leads to two critical points if $\ALPHA > \sqrt{k}$:
\begin{align*}
&\left( \ALPHA - \sqrt{\ALPHA^2-k}, \ALPHA + \sqrt{\ALPHA^2-k} \right) \\ 
&\left( \ALPHA + \sqrt{\ALPHA^2-k}, \ALPHA - \sqrt{\ALPHA^2-k} \right). \\ 
\end{align*}

On solving the cubic term explicitly, we observe that we get only one real solution. We arrange for this solution to be on the line $y = x$ by choosing $k = s^3/(2\ALPHA -s)$ (in which case the third critical point is $(s, s)$). On eliminating $\ALPHA, k$ from the above equations, we see that all critical points must lie on
\[
x y (x + y -s) = s^3.
\] A stability analysis shows that for any $s$, the critical point at $(s, s)$ is unstable, while the critical points $\left( \ALPHA - \sqrt{\ALPHA^2-k}, \ALPHA + \sqrt{\ALPHA^2-k} \right) $ and $\left( \ALPHA + \sqrt{\ALPHA^2-k}, \ALPHA - \sqrt{\ALPHA^2-k} \right)$ are stable.

Parameters can now be chosen by first choosing $s>0$, then $\ALPHA > s$ and finally $k = s^3/(2\ALPHA -s).$ We fix the parameter $\gamma = \ln(2)$ because we assume our units of time to be in terms of the protein half-life. Finally, we can solve for the parameter $\beta = 2\gamma \ALPHA.$

The parameters used in our study  are $s = 10, \gamma = \ln(2)$ and $\ALPHA = 15.8202.$ The phase space of the co-repressive toggle is mapped onto the RM as follows: Denote by $X$ and $Y$ the number of molecules of each protein type in the system. The region between the $y-$ axis and the $45^\circ$ line that passes half-way between the saddle and the stable point in the region $Y>X$ is mapped onto the state $H$. The corresponding region between the $x$-axis and the $45^\circ$ degree line in the $X>Y$ region is mapped into state $L$.  All trajectories are initialized at the saddle $(s, s)$, following which, a long transient is computed. All numerical estimates for mean residence times, failed transition probabilities, and stationary distributions are computed for $10^{4}$ transitions from state $H$ to $L$ (and back). 

\vspace*{1cm}

\begin{figure}[hb]
\includegraphics[width=6in]{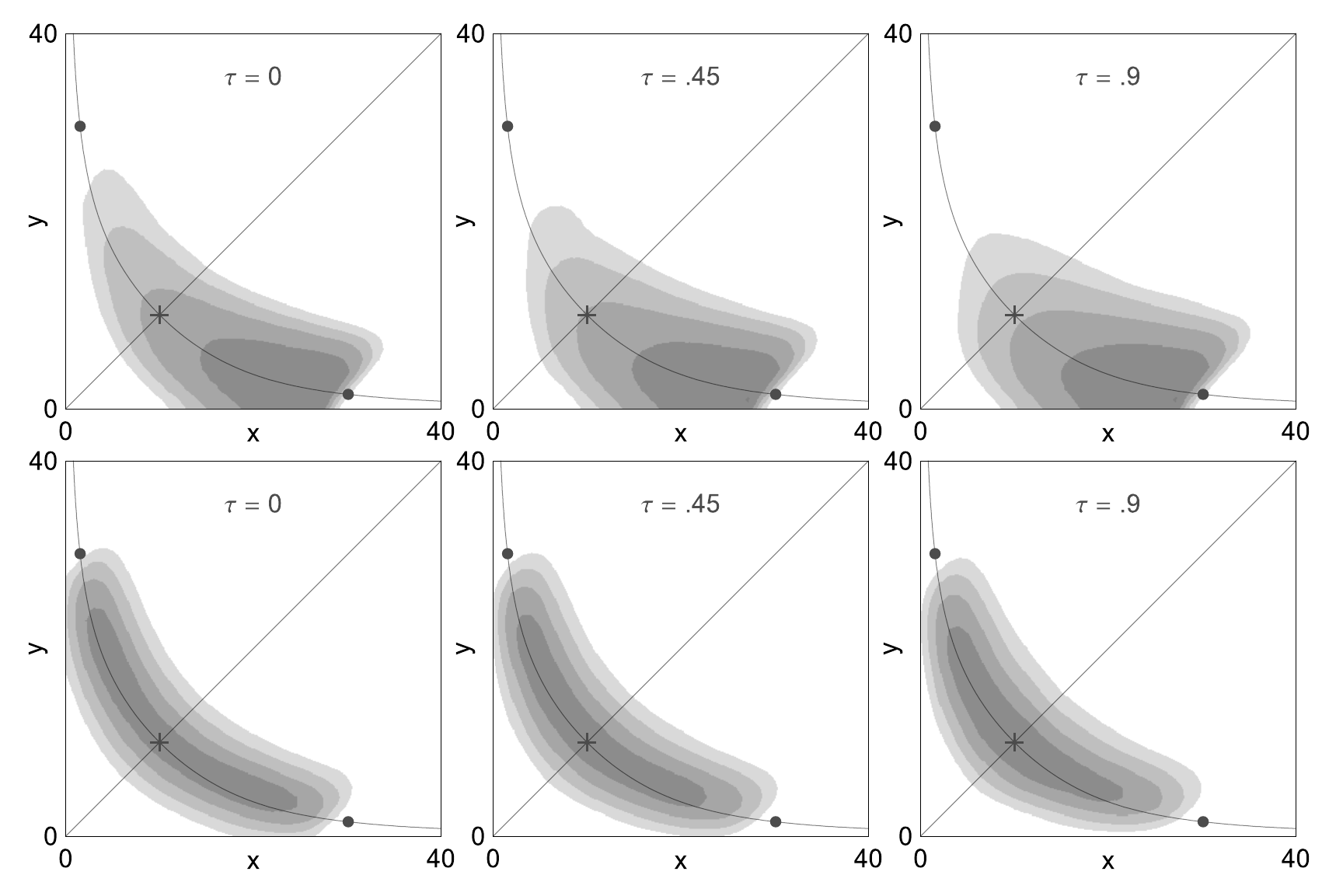}
\caption{
The top panels show the density corresponding to trajectories that make a failed transition attempt, starting in a neighborhood of the stable point in the region $X>Y$. With increasing $\tau$, the support of the density is more spread out. The bottom panels show the densities for the trajectories corresponding to successful transitions. Again, we observe that the support of the densities widens, although the effect is not as pronounced as in the case of failed transitions. 
}\label{fig:densities}
\end{figure}

\noindent\paragraph{Transition trajectories.} Delay also widens the distribution of paths that lead to failed transitions, as well as the distribution of those paths that correspond to successful transitions. The changes in the densities of the failed transition paths appear to be more sensitive to delay. Since the RM is not constructed using the specific features of any of the models, this effect cannot be explained using our reduction. We present in Figure-\ref{fig:densities} the densities of the paths that correspond to failed, and successful transitions, for the co-repressive toggle switch.

\end{document}